\documentclass[floats,prd,aps,amssymb,nofootinbib]{revtex4}
\usepackage{amssymb}
\usepackage{graphics}
\usepackage{amsmath}
\usepackage{amsfonts}
\usepackage{bm}
\usepackage{color}
\usepackage[]{latexsym}
\usepackage{amstext}
\usepackage{amssymb}
\usepackage{graphicx}
\usepackage{booktabs}

\usepackage{etoolbox} 

\newcommand{\be}{\begin{equation}}\newcommand{\ee}{\end{equation}}
\newcommand{\bea}{\begin{eqnarray}}\newcommand{\eea}{\end{eqnarray}}
\newcommand{\brr}{\begin{array}}\newcommand{\err}{\end{array}}
\newcommand{\bit}{\begin{itemize}}\newcommand{\eit}{\end{itemize}}
\newcommand{\ben}{\begin{enumerate}}\newcommand{\een}{\end{enumerate}}

\newcommand{\ba}{\begin{array}}
\newcommand{\ea}{\end{array}}

\def\lf{\left}

\def\non{\nonumber}\def\pa{\partial}

\def\ri{\right}
\def\al{\alpha}

\def\si{\sigma}

\def\1{{_{1}}}\def\2{{_{2}}}

\def\noHe0{:\;\!\!\;\!\!:H_e(0):\;\!\!\;\!\!:}
\def\noHm0{:\;\!\!\;\!\!:H_\mu(0):\;\!\!\;\!\!:}

\def\lf{\left}

\def\non{\nonumber}
\def\pa{\partial}

\def\ri{\right}

\def\al{\alpha}

\def\si{\sigma}

\def\1{{_{1}}}\def\2{{_{2}}}

\def\be{\begin{equation}}
\def\ee{\end{equation}}
\def\al{\alpha}
\def\bea{\begin{eqnarray}}
\def\eea{\end{eqnarray}}

\begin{document}
	
\title{Generalized uncertainty principle with maximal observable momentum and no minimal length indeterminacy}

\author{Luciano Petruzziello\footnote{lupetruzziello@unisa.it}$^{\hspace{0.3mm}1,2}$}
 
\affiliation
{\vspace{01mm}$^1$Dipartimento di Ingegneria, Universit\'a di Salerno, Via Giovanni Paolo II, 132 I-84084 Fisciano (SA), Italy.
\\ 
\vspace{0mm}
$^2$INFN, Sezione di Napoli, Gruppo collegato di Salerno, Italy.}

\begin{abstract}
We present a novel generalization of the Heisenberg uncertainty principle which introduces the existence of a maximal observable momentum and at the same time does not entail a minimal indeterminacy in position. The above result is an \emph{exact} generalized uncertainty principle (GUP), valid at all energy scales. For small values of the deformation parameter $\beta$, our ansatz is consistent with the usual expression for GUP borrowed from string theory, doubly special relativity and other quantum gravity candidates that provide $\beta$ with a negative sign. As a preliminary analysis, we study the implications of this new model on some quantum mechanical applications and on the black hole thermodynamics.   
\end{abstract}

 \vskip -1.0 truecm
\maketitle

\section{Introduction}

Among the most striking predictions of several candidates of quantum gravity it is possible to recognize the existence of a minimal length at Planck scale~\cite{string}. Such an awareness necessarily dictates a modification of the usual Heisenberg uncertainty principle (HUP), according to which sufficiently energetic probes may investigate arbitrarily small spatial resolutions.
The first and most immediate generalization of HUP that accounts for the presence of a minimal length implies the inclusion of a momentum-dependent term, which typically lets us cast the product $\Delta x\Delta p$ in the form~\cite{kmm,scard}
\be\label{kmmgup}
\Delta x\Delta p\geq \frac{\hbar}{2}\lf(1+\beta{\Delta p^2}\ri)\equiv\frac{\hbar}{2}\lf(1+\beta_0\frac{\Delta p^2}{m_p^2c^2}\ri),
\ee
with $m_p$ being the Planck mass and ($\beta_0$) $\beta$ the (dimensionless) deformation parameter. Henceforth, we will address the above expression as the Kempf-Mangano-Mann (KMM) generalized uncertainty principle (GUP). The magnitude of $\beta_0$ is assumed to be of order unity in string theory~\cite{string}, and analogous outcomes have been achieved also in other physical scenarios. For instance, we recall that, from a straightforward analysis of the quantum corrections to the Newtonian potential, one can derive $\beta_0=82\pi/5$~\cite{lamb}, whereas by studying the deformed Unruh temperature in the maximal acceleration framework one would obtain $\beta_0=8\pi^2/9$~\cite{maxacc}. Similar results have been found out in the most disparate contexts, ranging from the non-commutative Schwarzschild geometry~\cite{noncom2} to the corpuscular gravity description of black holes~\cite{set}, thus further corroborating the predictions of string theory. It is worth observing that the KMM GUP has been extensively investigated in many theoretical settings, as the vast literature on this topic suggests~\cite{lamb,maxacc,noncom2,set,noui,qft,qftbis,iorio,qftter,czech,noncom,bis,ter,ter2,quar,ong,ot,das,nonext,casascard,bek}.

Along the line of the KMM GUP, several attempts have been performed to find a completion of Eq.~\eqref{kmmgup} which is valid at all energy scales. Strictly speaking, the generalized version of HUP presented in Eq.~\eqref{kmmgup} is believed to represent only the leading term of the expansion in the small parameter $\sqrt{\beta}\Delta p$ of a higher-order GUP. In this regard, the first proposal that broadens the validity domain of the KMM GUP is due to Nouicer~\cite{noui2}, which preserves the existence of the minimal length uncertainty and the arbitrarily large momentum. The latter aspect is superseded in the approach moved forward by Pedram~\cite{pedram}, in which the form of GUP naturally induces the introduction of a UV cutoff at the Planck scale for the momentum, a feature shared also with the model conceived by Hassanabadi and Chung (HC)~\cite{hc} and concurrently by Shababi and Chung (SC)~\cite{sc}. 
In order to properly compare all the above uncertainty relations, we summarize their main characteristics in Table~\ref{table} at the end of the Section.

In view of the picture outlined so far, the deformation parameter is considered to be a positive quantity. However, a significant number of papers (i.e. Refs.~\cite{noncom2,ter2,ong,set} and references therein) comply with the vision of a negative value for $\beta$, thereby allowing for a classical regime at the Planck scale, since for $\beta<0$ we can deduce from Eq.~\eqref{kmmgup} that if $\Delta p\simeq m_pc$, then $\Delta x\Delta p\geq0$. A similar circumstance is not entirely unprecedented, as it has already been discussed in the framework of doubly special relativity~\cite{dsr} as well as in the deterministic interpretation of quantum mechanics~\cite{thooft}. Despite this, a higher-order GUP which possesses the aforementioned property and at the same time provides for a natural cutoff for the momentum is nowhere to be found. The purpose of the present paper is precisely to tackle the previous issue by trying to extend HUP in such a way not to include a minimal length uncertainty \emph{and} to have a maximal observable momentum. By virtue of the affinity with doubly special relativity, this uncertainty relation would be applicable to a wide variety of problems and hence it may potentially be viewed as a further step towards the establishment of a physically plausible quantum theory of gravity. 

In order to fulfill our intent, we organize the manuscript as follows: in Sec.~II we introduce our higher-order GUP and show its main aspects, among which we include the functional analysis of the position operator and the harmonic oscillator. Section~III is devoted to the employment of the aforesaid formalism for a thorough study of black hole thermodynamics, whilst Sec.~IV contains concluding remarks and future perspectives.

\begin{table}[ht]
\caption{Comparison of higher-order generalized uncertainty relations.} 
\centering 
\begin{tabular}{c c c c} 
\hline\hline 
Uncertainty principle &\hspace{4mm} Shape for $\langle p\rangle=0$ &\hspace{4mm} Minimal length uncertainty &\hspace{4mm} Maximal observable momentum \\ [0.5ex] 

\hline \vspace{1mm}

Heisenberg &\hspace{4mm} $\Delta x\Delta p\geq\frac{\hbar}{2}$ &\hspace{4mm} $\times$ &\hspace{4mm} $\times$ \\ \vspace{2mm}
KMM &\hspace{4mm} $\Delta x\Delta p\geq\frac{\hbar}{2}\lf(1+\beta\Delta p^2\ri)$ &\hspace{4mm} \checkmark &\hspace{4mm} $\times$  \\ \vspace{2mm}
Nouicer &\hspace{4mm} $\Delta x\Delta p\geq\frac{\hbar}{2}e^{\beta \Delta p^2}$ &\hspace{4mm} \checkmark &\hspace{4mm} $\times$ \\ \vspace{2mm}
Pedram &\hspace{4mm} $\Delta x\Delta p\geq\frac{\hbar}{2}\frac{1}{1-\beta\Delta p^2}$ &\hspace{4mm} \checkmark &\hspace{4mm} \checkmark \\ \vspace{2mm}
HC &\hspace{4mm} $\Delta x\Delta p\geq\frac{\hbar}{2}\lf(-\beta\Delta p+\frac{1}{1-\beta\Delta p}\ri)$ &\hspace{4mm} \checkmark &\hspace{4mm} \checkmark \\ \vspace{2mm}
SC &\hspace{4mm} $\Delta x\Delta p\geq\frac{\hbar}{2}\frac{2\beta \Delta p^2\sqrt{1+4\beta \Delta p^2}}{\sqrt{1+4\beta \Delta p^2}-1}$ &\hspace{4mm} \checkmark &\hspace{4mm} \checkmark \\ [1ex] 
\hline\hline
\end{tabular}
\label{table} 
\end{table}

\section{Higher-order GUP}

Before the beginning, it is opportune to observe that the generalized uncertainty principles exhibited in Table~I stem from a suitable modification of the canonical commutation relation between the momentum and position operators. For example, the KMM GUP expressed in Eq.~\eqref{kmmgup} can be deduced from the following commutator:
\be\label{kmmcomm}
\bigl[\hat{x},\hat{p}\bigr]=i\hbar\lf(1\pm\lf|\beta\ri| \hat{p}^2\ri), 
\ee
where the sign of $\beta$ can be taken either positive or negative by virtue of the aforementioned discussions\footnote{Clearly, the expression~\eqref{kmmgup} is recovered for the selection of the positive sign in Eq.~\eqref{kmmcomm}.}.

In compliance with our purpose, we now seek the appropriate formula for the canonical commutator that allows for the existence of a maximal momentum and reduces to the previous equation with the minus sign in the regime $\sqrt{\beta}\,{p}\ll1$, so as to exclude a minimal uncertainty in position. To simultaneously satisfy the aforesaid requirements, we introduce our proposal for GUP by postulating that the extension of Eq.~\eqref{kmmcomm} with the choice $-|\beta|$ is
\be\label{ourcomm}
\bigl[\hat{x},\hat{p}\bigr]=i\hbar\,\sqrt{1-2\beta \hat{p}^2}\,,
\ee
where we have removed the absolute value of the deformation parameter to ease the notation. Equation~\eqref{ourcomm} immediately conveys the concept that the momentum cannot exceed the value $p_{max}=1/\sqrt{2\beta}$, otherwise the whole expression loses its physical meaning. Furthermore, for $p=\pm p_{max}$ it is evident that the position and momentum operators commute, just like in the classical scenario. These findings definitely let us identify $p_{max}$ with the desired high-energy cutoff at the Planck scale.

\begin{figure}[ht]
\centering
  \includegraphics[width=18cm]{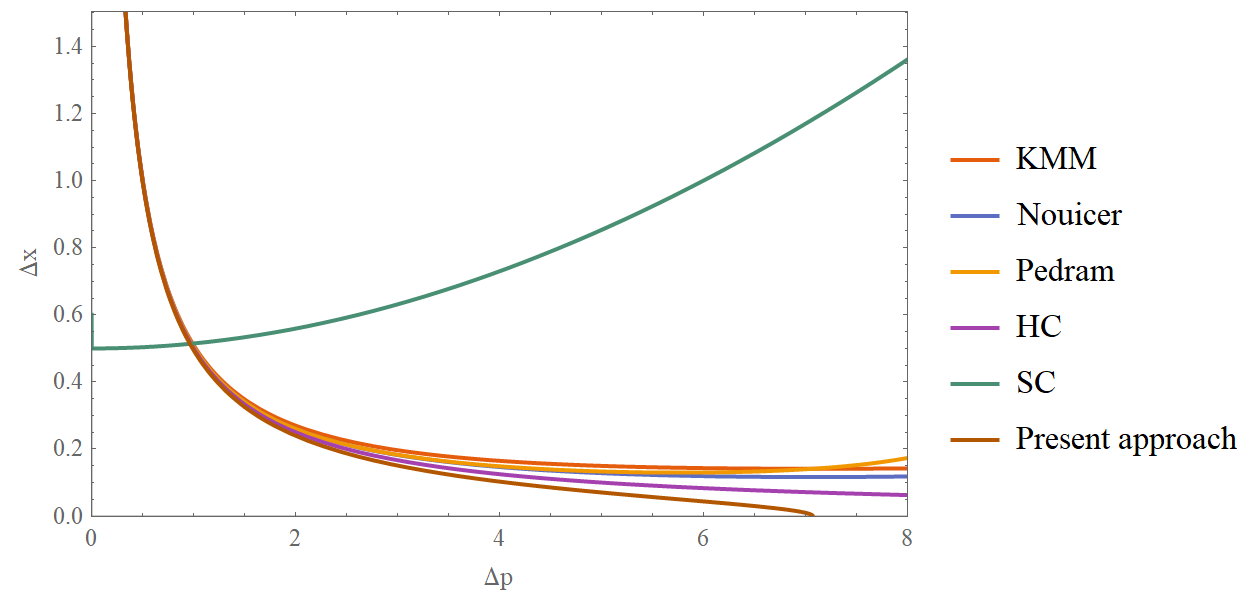}
  \caption{The plot of $\Delta x$ versus $\Delta p$ for several GUPs. We have conveniently chosen $\beta=0.01$ and $\hbar=1$ so as to simultaneously compare all the uncertainty relations.}
  \label{fig}
\end{figure}

To derive the uncertainty relations, we start from~\eqref{ourcomm} and note that
\bea\non
\Delta x\Delta p&\geq&\frac{\hbar}{2}\lf\langle\sqrt{1-2\beta \hat{p}^2}\ri\rangle\geq\frac{\hbar}{2}\sum_{n=0}^\infty\frac{\beta^n(2n)!}{2^n(1-2n)(n!)^2}\langle \hat{p}^{2n}\rangle\\[2mm]\non
&\geq&\frac{\hbar}{2}\sum_{n=0}^\infty\frac{\beta^n(2n)!}{2^n(1-2n)(n!)^2}\langle \hat{p}^2\rangle^n\geq\frac{\hbar}{2}\sum_{n=0}^\infty\frac{\beta^n(2n)!}{2^n(1-2n)(n!)^2}\lf(\Delta p^2+\langle \hat{p}\rangle^2\ri)^n\\[2mm]\label{ourgup}
&\geq&\frac{\hbar}{2}\sqrt{1-2\beta\lf(\Delta p^2+\langle \hat{p}\rangle^2\ri)}\,,
\eea
where we have made use of the property $\langle \hat{p}^{2n}\rangle\geq\langle \hat{p}^2\rangle^n$ in conformity with Refs.~\cite{noui2,pedram,hc,sc}. For the sake of simplicity, we will restrict the attention to mirror-symmetric states, for which $\langle \hat{p}\rangle=0$. As already anticipated, for $\sqrt{\beta}\Delta p\ll1$ we recover the usual KMM GUP~\eqref{kmmgup} with the negative sign for the deformation parameter, in line with Refs.~\cite{noncom2,ter2,ong,set}. To better emphasize the features of this new higher-order GUP, in Fig.~\ref{fig} we plot all the uncertainty relations contained in Table~\ref{table} together with Eq.~\eqref{ourgup} when the equalities are saturated.  

\noindent
To grasp a further insight on the present form of GUP, we observe that a viable representation for the position and momentum operator can be achieved in the momentum space representation, where 
\be\label{rep}
\hat{p}\,\psi(p)=p\,\psi(p)\,, \qquad \hat{x}\,\psi(p)=i\hbar\sqrt{1-2\beta p^2}\,\pa_p\psi(p)\,.
\ee
According to the above scheme, the completeness relation and the scalar product can be respectively defined as
\be\label{qmgup}
\langle p|p'\rangle=\sqrt{1-2\beta p^2}\,\delta(p-p')\,, \qquad
\langle\psi|\chi\rangle=\int^{1/\sqrt{2\beta}}_{-1/\sqrt{2\beta}}\frac{\psi^*(p)\chi(p)}{\sqrt{1-2\beta p^2}}\,dp\,,
\ee
where in the second equation we have correctly taken into account the maximal allowed value for the momentum.

\subsection*{Functional analysis of the position operator}

In light of Eq.~\eqref{rep}, we can study the eigenvalue problem for the position operator, which amounts to solve the differential equation
\be\label{deq}
i\hbar\sqrt{1-2\beta p^2}\,\pa_p\psi_\lambda(p)=\lambda\,\psi_\lambda(p)\,,
\ee
whose solution is represented by
\be\label{solution}
\psi_\lambda(p)=k\,\exp\lf[-\frac{i\,\lambda\,\mathrm{arcsin}\lf(\sqrt{2\beta}\,p\ri)}{\sqrt{2\beta}\,\hbar}\ri]\,,
\ee
which reduces to the standard expression for the quantum mechanical plane waves as long as $\beta\to0$. To find the normalization factor $k$, we have to impose the condition
\be\label{norm}
\langle\psi_\lambda|\psi_\lambda\rangle=\int^{1/\sqrt{2\beta}}_{-1/\sqrt{2\beta}}\frac{|k|^2}{\sqrt{1-2\beta p^2}}\,dp=1\,,
\ee
which is realized (up to an overall phase factor) for 
\be\label{k}
k=\sqrt{\frac{\sqrt{2\beta}}{\pi}}\,.
\ee
For the sake of completeness, we show that also for this form of GUP the position eigenstates are no longer orthogonal, since
\be\label{sp}
\langle\psi_\lambda|\psi_{\lambda'}\rangle=\frac{\sqrt{2\beta}}{\pi}\int^{1/\sqrt{2\beta}}_{-1/\sqrt{2\beta}}\frac{\exp\lf[\frac{i(\lambda-\lambda')\mathrm{arcsin}\lf(\sqrt{2\beta}\,p\ri)}{\sqrt{2\beta}\,\hbar}\ri]}{\sqrt{1-2\beta p^2}}\,dp=\frac{2\sqrt{2\beta}\,\hbar}{\pi(\lambda-\lambda')}\sin\lf[\frac{\pi(\lambda-\lambda')}{2\sqrt{2\beta}\,\hbar}\ri]\,.
\ee
Such a feature is in agreement with the previous models of generalized uncertainty principle~\cite{kmm,noui2,pedram,hc,sc}. However, differently from the other approaches, for the present case it is not necessary to proceed with the description of maximally localized states as done for the other GUPs. As a matter of fact, the absence of a lower bound for $\Delta x$ prevents us from recognizing states $|\psi_{ml}\rangle$ on which the position uncertainty equals the minimal length indeterminacy predicted by the theory. 

\subsection*{Harmonic oscillator}

As a final preliminary application, it is constructive to analyze the one-dimensional harmonic oscillator, whose Hamiltonian is 
\be\label{harmonic}
\hat{H}=\frac{\hat{p}^2}{2m}+\frac{m\omega^2\hat{x}^2}{2}\,,
\ee
with $m$ being the mass of the particle and $\omega$ the angular frequency of the oscillator. By resorting to Eq.~\eqref{rep} and assuming the absence of an explicit dependence on time, the stationary state Schr\"odinger equation in the momentum space representation becomes
\be\label{stat}
\frac{p^2}{2m}\psi-\frac{\hbar^2 m\omega^2}{2}\sqrt{1-2\beta p^2}\,\frac{d}{dp}\lf(\sqrt{1-2\beta p^2}\,\frac{d\psi}{dp}\ri)=E\psi\,,
\ee
where we recall that $p\in\lf[-1/\sqrt{2\beta},1/\sqrt{2\beta}\ri]$. After several algebraic manipulations, the above expression can be reformulated in terms of the dimensionless parameter $\kappa=\sqrt{\beta}\,p$, which ranges in the interval $\kappa\in\lf[-1/\sqrt{2},1/\sqrt{2}\ri]$; hence, we have 
\be\label{stat2}
\frac{d^2\psi}{d\kappa^2}-\frac{2\kappa}{1-2\kappa^2}\frac{d\psi}{d\kappa}+\frac{2\beta mE-\kappa^2}{\beta^2\hbar^2 m^2\omega^2(1-2\kappa^2)}\psi=0\,.
\ee
To further simplify the differential equation, we perform the following change of variable:
\be\label{change}
y=\frac{\mathrm{arcsin}\lf(\sqrt{2}\,\kappa\ri)}{\sqrt{2}}\,, \qquad y\in\lf[-\frac{\pi}{2\sqrt{2}},\frac{\pi}{2\sqrt{2}}\ri],
\ee
so as to remove the first-order derivative and remain with
\be\label{stat3}
\frac{d^2\psi}{dy^2}+V\psi=0\,, \qquad 
\ee
where
\be\label{pot}
V\equiv V(y)=\frac{4\beta mE-\sin^2\lf(\sqrt{2}\,y\ri)}{2\beta^2\hbar^2 m^2\omega^2}
\ee
is the effective potential. The differential equation~\eqref{stat3} is commonly encountered in literature, and it goes by the name of ``quantum pendulum''~\cite{pendulum}, which properly describes molecular motion and chaotic dynamics. The most general solution of the quantum pendulum can be written as a linear superposition of non-periodic Mathieu functions~\cite{grad}, but the values of the discrete energy levels that arise from such a study cannot be found analytically~\cite{oscillator}. Nonetheless, the problem can still be tackled numerically; in so doing, one can rebuild the energy spectrum up to a significant number of terms\footnote{Keep in mind that we do not need to evaluate all the arbitrarily large number of contributions. Indeed, due to the fact that the momentum is bounded, the energy spectrum for the harmonic oscillator is finite.}. For more technical details, we remand the interested reader to Ref.~\cite{pendulum}.

\section{Black hole thermodynamics}

\begin{figure}[ht]
\centering
  \includegraphics[width=18cm]{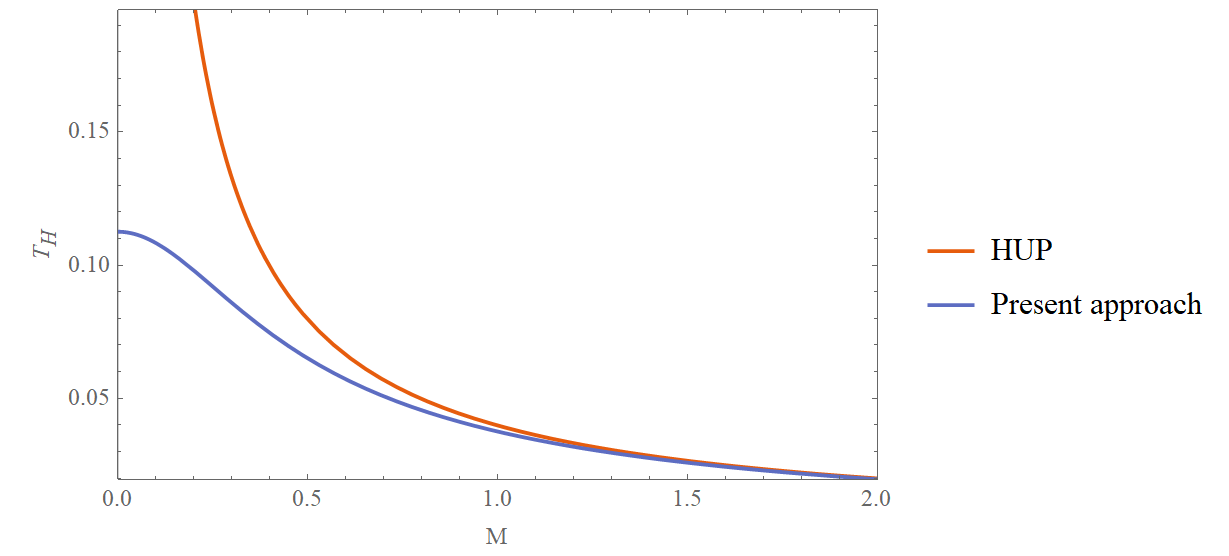}
  \caption{The plot of the black hole radiation temperature versus the mass $M$ for the standard result and for its deformation due to the present form of GUP. For simplicity, we have set $\hbar=c=k_B=G=\beta=1$.}
  \label{fig2}
\end{figure}

In this Section, we closely follow the heuristic approaches adopted in Refs.~\cite{scard,bis,set} to describe the main modifications that the settlement of the uncertainty relation~\eqref{ourgup} entails at the level of the standard black hole thermodynamics. To this aim, let us work with a spherically symmetric black hole of mass $M$; consequently, the metric for a similar setting in spherical coordinates is given by
\be\label{metric}
ds^2=\lf(1-\frac{r_s}{r}\ri)c^2dt^2-\lf(1-\frac{r_s}{r}\ri)^{-1}dr^2-r^2\lf(d\theta^2+\sin^2\theta d\varphi^2\ri)\,,
\ee
with $r_s=2GM/c^2$ being the Schwarzschild radius. At this point, let us focus the attention on a photon which has been emitted by the black hole and it is just outside the event horizon. We can safely consider that the position uncertainty of such particle is proportional to $r_s$, namely $\Delta x\simeq r_s$. By virtue of Eq.~\eqref{ourgup}, we then obtain
\be\label{dep}
\Delta p\simeq\frac{\hbar c^2}{4GM}\sqrt{1-2\beta\Delta p^2}\,.
\ee
Since $\Delta p$ can be associated to the average energy of the emitted photon as $E\simeq c\Delta p$, the previous expression can yield the modified Hawking temperature $T_H$ by relying on the equipartition theorem, which allows us to identify $E$ with the temperature of the photon radiation, i.e. $E\simeq k_BT_H$. In so doing, the formula for $T_H$ is
\be\label{hawtemp}
T_H=\frac{\al\hbar c^3}{4k_BGM}\frac{1}{\sqrt{1+\frac{\beta\hbar^2}{2r_s^2}}}\,,
\ee
with $\al$ being a calibration factor that incorporates all the approximations made so far and that should be fixed consistently with the semi-classical result obtained in the limit $\beta\to0$; this is realized when $\al=1/2\pi$. The behavior of $T_H$ as a function of the black hole mass is exhibited in Fig.~\ref{fig2}.

\noindent
From the above figure, it is possible to deduce that, at the end of the evaporation process, the temperature $T_H$ does not diverge, but instead will smoothly reach a constant value which is independent of the initial mass of the black hole. Interestingly, this universal quantity is proportional to the Planck temperature $T_p$, and more precisely it is equal to
\be\label{temp}
T_{max}\simeq\sqrt{\frac{\hbar c^5}{\beta_0k_B^2G}}=\frac{T_p}{\sqrt{\beta_0}}\simeq\frac{10^{32}}{\sqrt{\beta_0}}\,K\,.
\ee
If we want the perfect match $T_{max}=T_p$, we should require $\beta_0\simeq\mathcal{O}(1)$, which is in agreement with the claims revolving around the deformation parameter.

\begin{figure}[ht]
\centering
  \includegraphics[width=18cm]{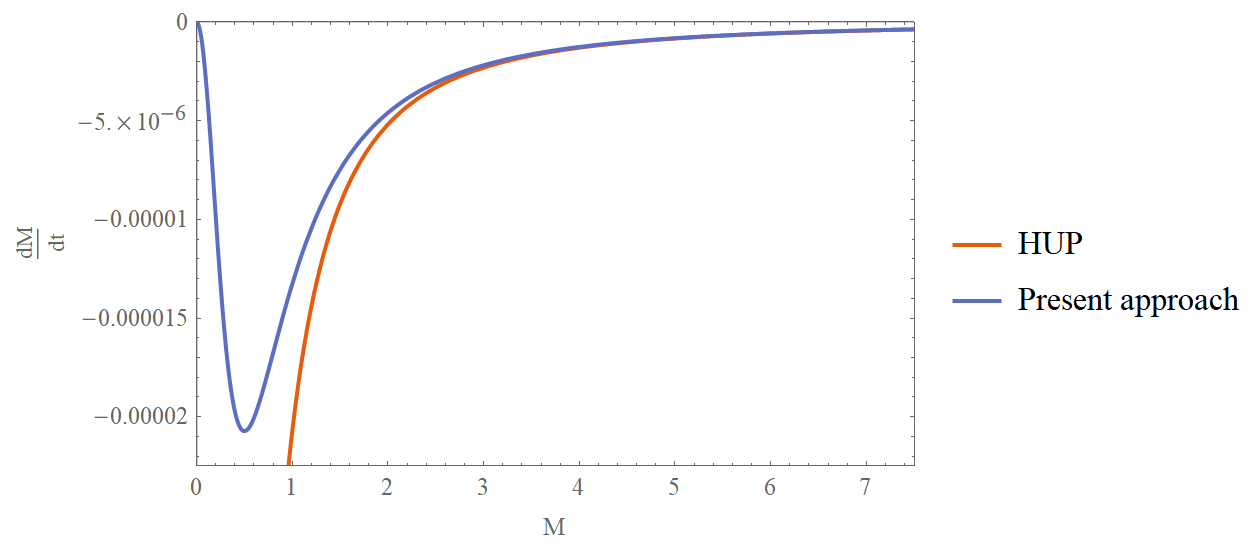}
  \caption{The plot of the black hole emission rate versus the mass $M$ for the standard result and for its deformation due to the present form of GUP. As before, we have set $\hbar=c=k_B=G=\beta=1$.}
  \label{fig3}
\end{figure}

As a further step, the formula~\eqref{hawtemp} is capable of unveiling the ultimate fate of the black hole predicted by the current model of GUP. As a matter of fact, as firstly pointed out in Ref.~\cite{bis} for the case $\beta>0$, the physical impossibility for $M$ to assume arbitrarily small values is a signature which indicates the emergence of an inert remnant with finite size at the end of the evaporation process. Apparently, a similar circumstance does not occur for our GUP, since the deformed Hawking temperature is well-defined for any value of $M$, even when $M\to0$. Such a feature is shared also with the evaporation rate, which can be readily achieved by exploiting the Stefan-Boltzmann law for an estimation of the radiated power $P$, that is
\be\label{sb}
P=\si A_sT_H^4\,,
\ee
where $\si=\pi^2k_B^4/(60\hbar^3c^2)$ is the Stefan-Boltzmann constant and $A_s=4\pi r_s^2$ is the surface area of the black hole, which here is regarded as a blackbody. Therefore, the decrease of mass over time is simply given by
\be\label{erate}
\frac{dM}{dt}=-\frac{P}{c^2}=-\frac{\hbar c^4}{15360\pi G^2M^2}{\lf(1+\frac{\beta\hbar^2}{2r_s^2}\ri)^{-2}}\,,
\ee
which clearly furnishes the usual result for $\beta\to0$. The interesting aspect of Eq.~\eqref{erate} is the avoidance of the negative divergence of $dM/dt$, which on the contrary vanishes at the end of the evaporation process, as one may have correctly expected (see Fig.~\ref{fig3} for further details). Before that, however, the emission rate reaches a minimum as $M$ approaches the Planck mass, namely 
\be\label{min}
M_{min}\simeq{\sqrt{\beta_0}}\,m_p\,,
\ee 
after which the function steadily tends to zero. Such a behavior could in principle be ascribable to quantum gravitational effects, as they are deemed to be dominant at the Planck scale.

\noindent
Starting from Eq.~\eqref{erate}, it is possible to compute also the evaporation time $t_{ev}$; as a matter of fact, we have
\be\label{tev}
t_{ev}=-\frac{15360\pi G^2}{\hbar c^4}\lim_{\varepsilon\to0}\int_M^\varepsilon x^2\lf(1+\frac{\beta\hbar^2c^4}{8G^2x^2}\ri)^2dx\,,
\ee
that contains the missing information to unravel the destiny of the black hole.
Indeed, the above integral does not converge, which means that the completion of the black hole evaporation would require an infinite amount of time to be finalized, thus always leaving a remnant behind as a product of the long-term process. This implies that $T_H$ and $dM/dt$ would only asymptotically reach the values $T_{max}$ and zero, respectively. The aforesaid claim is not entirely new, as it has been already encountered in the interplay between black hole physics and generalized uncertainty principle with negative deformation parameter~\cite{yong}.

\begin{figure}[ht]
\centering
  \includegraphics[width=18cm]{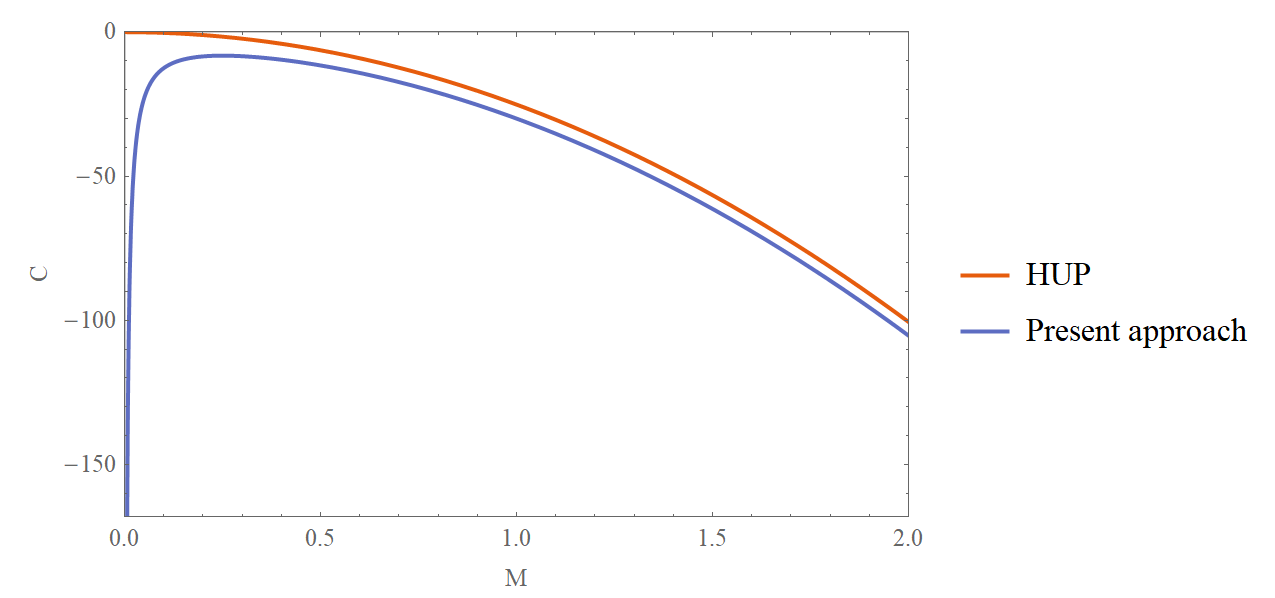}
  \caption{The plot of the heat capacity versus the mass $M$ for the standard result and for its deformation due to the present form of GUP. Again, we have set $\hbar=c=k_B=G=\beta=1$.}
  \label{fig4}
\end{figure}

To check whether the black hole eventually becomes thermodynamically inert at some point, we need to calculate the heat capacity, defined as $C=c^2dM/dT$. Such a computation yields
\be\label{heat}
C=-\frac{8\pi k_BGM^2}{\hbar c}\lf({1+\frac{\beta\hbar^2}{2r_s^2}}\ri)^{\frac{3}{2}}\,,
\ee
which confirms the fact that $C$ is a negative quantity for each value of $M$, an aspect already present in the standard framework.
Despite this, the peculiarity of Eq.~\eqref{heat} is that it leads to an asymptotic divergence of the heat capacity when $M\to0$, as it can be seen in Fig.~\ref{fig4}. A similar phenomenon entails that the black hole would eventually undergo a metastable regime which concretely resembles a quantum phase transition. In this sense, our statement is analogous to the corpuscular gravity description proposed by Dvali~\cite{set,dvali}. 

\noindent
Remarkably, the maximum of $C$ is achieved for the same value of the mass $M$ at which there is a minimum in the evaporation rate (see Eq.~\eqref{min}) and as $dM/dt$ gradually vanishes, the heat capacity blows up\footnote{This occurrence is harmless, since the black hole requires an infinite time to fully disappear, and hence the divergence is realized only asymptotically.}. Consequently, it is licit to assume that, at the Planck scale, the remnant dramatically slows down its mass decrement and simultaneously enhances the thermodynamical interaction with the surrounding environment while being on the verge of a phase-transition-like status. As already mentioned before, a more rigorous explanation of such a tendency can only be attained with a consistent model of quantum gravity.

As a final remark, we can evaluate the black hole entropy. For this purpose, we resort to the identity $T_HdS=c^2dM$, as seen in Ref.~\cite{bis}. The outcome of this procedure gives
\be\label{entropy}
S=c^2\int\frac{dM}{T_H}=\sqrt{2}\,\pi k_B\xi^2\sqrt{8+\frac{\beta_0}{\xi^2}}+\frac{\pi\beta_0k_B}{4}\ln\lf[\xi^2\lf(4+\sqrt{16+\frac{2\beta_0}{\xi^2}}\ri)^2\ri]=S_{BH}\sqrt{1+\frac{\beta_0}{8\xi^2}}+\delta S\,,
\ee
with $\xi=M/m_p$ and $S_{BH}=k_BA_sc^3/(4G\hbar)$. Note that in the previous expression we can recognize a logarithmic correction to the standard Bekenstein-Hawking entropy which depends upon the area of the black hole, in accordance with several works appeared in literature~\cite{bis,log}. Nonetheless, the sign of such a contribution is positive, which unambiguously indicates that $S>S_{BH}$, thereby apparently violating the Bekenstein bound~\cite{bekenstein}. On the other hand, a more accurate analysis shows that our result is not paradoxical at all, since the very presence of GUP induces a generalization of the fundamental inequality discovered by Bekenstein, as proved in Ref.~\cite{bek}.

\section{Conclusions}

In this paper, we have discussed for the first time a higher-order GUP which predicts a maximal observable momentum and no minimal length uncertainty. Specifically, the result $\Delta x=0$ is achieved for $\Delta p=\pm1/\sqrt{2\beta}$, that is the maximally allowed quantity for the momentum, as conveyed by Eqs.~\eqref{ourcomm} and~\eqref{ourgup}. For $\sqrt{\beta}\,p\ll1$, this novel form of GUP is in accordance with the well-known expression~\eqref{kmmgup} with a negative value for the deformation parameter, which has been the subject of intense investigation in recent literature (see Refs.~\cite{noncom2,ter2,ong,set} and references therein). It is worth pointing out once again that the present extension of the uncertainty relation is significantly different from all the other higher-order GUPs~\cite{noui2,pedram,hc,sc}, as it can be deduced from the study of the applications and their results. For the above reasons, our approach is tailor-made to fit in the formalism of quantum gravity proposals such as doubly special relativity~\cite{dsr} and to explain the existence of a classical regime at the Planck scale, as suggested in the aforementioned references but also in the approach to quantum mechanics due to 't Hooft~\cite{thooft}. 

To preliminarily discuss the implications of the current model, we have firstly studied the functional behavior of the position operator, from which we have understood that position eigenvalues are no longer orthogonal, as seen in Refs.~\cite{kmm,noui2,pedram,hc,sc}. After that, we have tackled the topic of the harmonic oscillator and showed how it can be related to the problem of the quantum oscillator, thereby stressing that it possesses a discrete energy spectrum which is bounded from above due to the restrictions imposed by GUP on momentum. Subsequently, we have centered the discussion around the black hole thermodynamics and proved that, according to the present picture, such astrophysical objects invest an infinite amount of time to fully evaporate, leaving a metastable remnant behind that prominently interacts after reaching the Planck scale, where the temporal evolution of the mass decrement is sharply dampened. As a further observation, we have derived a logarithmic correction to the usual black hole entropy $S_{BH}$ that has a positive sign; however, such an occurrence is not in contradiction with the Bekenstein bound, as seen in Ref.~\cite{bek}.

Despite the accomplishment of these findings, there is still a plethora of problems and scenarios for which the higher-order GUP~\eqref{ourgup} can be employed. Therefore, more work along this line is inevitably required.

\end{document}